\documentclass[12pt,a4paper,final]{iopart}

\usepackage{iopams}  
\usepackage{graphicx}
\usepackage{verbatim}
\usepackage{textcomp,bbm}
\usepackage[breaklinks=true,colorlinks=true,linkcolor=blue,urlcolor=blue,citecolor=blue]{hyperref}
\usepackage{cite}

\newcommand{\F}{\mathcal{F}}

\newcommand{\um}{\,\textmu m}

\begin{document}

\title{Transverse-mode coupling effects in scanning cavity microscopy}

\author{Julia Benedikter$^{1,2}$, Thea Moosmayer$^{2}$, Matthias Mader$^{1,3}$, Thomas H\"ummer$^{1,3}$, %Theodor W. H\"ansch$^{1,3}$, 
	David Hunger$^{2}$}
\address{$^1$Fakult{\"a}t f{\"u}r Physik, Ludwig-Maximilians-Universit{\"a}t, Schellingstra{\ss}e~4, 80799~M{\"u}nchen, Germany}
\address{$^2$Karlsruher Institut f\"ur Technologie, Physikalisches Institut, Wolfgang-Gaede-Str. 1, 76131 Karlsruhe, Germany}
\address{$^3$Max-Planck-Institut f{\"u}r Quantenoptik,  Hans-Kopfermann-Str.~1, 85748~Garching, Germany}
\ead{david.hunger@kit.edu}

\begin{abstract}
Tunable open-access Fabry-Pérot microcavities enable the combination of cavity enhancement with high resolution imaging. To assess the limits of this technique originating from background variations, we perform high-finesse scanning cavity microscopy of pristine planar mirrors. We observe spatially localized features of strong cavity transmission reduction for certain cavity mode orders, and periodic background patterns with high spatial frequency. We show in detailed measurements that the localized structures originate from resonant transverse-mode coupling and arise from the topography of the planar mirror surface, in particular its local curvature and gradient. We further examine the background patterns and find that they derive from non-resonant mode coupling, and we attribute it to the micro roughness of the mirror. Our measurements and analysis elucidate the impact of imperfect mirrors and reveal the influence of their microscopic topography. This is crucial for the interpretation of scanning cavity images, and could provide relevant insight for precision applications such as gravitational wave detectors, laser gyroscopes, and reference cavities.
\end{abstract}

%\pacs{42.25.Fx, 42.15.Eq, 42.50.Wk, 42.60.Da}
\vspace{2pc}
\maketitle

\section{Introduction}
Optical microcavities are a powerful tool to enhance light-matter interactions for a variety of applications \cite{Vahala03}. With its capability to combine cavity enhancement with optical microscopy, scanning cavity microscopy (SCM) is emerging as a powerful technique for single-particle sensing and spectroscopy of heterogeneous nanomaterials \cite{Mader15,Kelkar15,Huemmer16}, and for the realization of efficient light-matter interfaces for solid-state quantum emitters \cite{Toninelli10,Kaupp13,Albrecht13,Greuter14,Janitz15,Johnson15,Kaupp16,Benedikter17,Riedel17,Bogdanovic17,Wang17,Casabone18}. The basic functionality derives from a fully tunable, open-access Fabry-Perot cavity with micron-scale mode size, which is typically realized by the combination of a micro-machined concave and a macroscopic planar mirror, the latter carrying the sample to be studied. One of the two mirrors can be raster-scanned laterally to obtain spatially resolved measurements. For the example of sensing and spectroscopy, small changes in the resonant cavity transmission are recorded to obtain information about sample-induced scattering and absorption \cite{Mader15,Kelkar15,Huemmer16}. In consequence, any background variations in transmission originating from the bare cavity alone need to be as small as possible to enable the resolution of small signals.

In different experiments \cite{Mader15,Huemmer16,Benedikter17} using various mirror coatings, cavity geometries, and wavelengths, we consistently observe two classes of artefacts in SCM images: spatially localized contour lines with deteriorated cavity performance for certain longitudinal mode orders, and weak periodic background patterns. 

To understand and quantify the effects, we perform extensive SCM measurements of pristine planar dielectric mirrors with high reflectivity. We find that both types of artefacts are related to transverse mode coupling, which is present due to imperfect mirror shapes \cite{Benedikter15}. 

Different techniques have been developed to produce concave, near-spherical profiles as micro-mirror substrates, including CO$_2$ laser machining \cite{Hunger12,Greuter14,Takahashi14,Muller09,Uphoff15}, chemical etching \cite{Trupke05,Biedermann10}, focused ion beam milling \cite{Dolan10,Albrecht14}, and thermal reflow \cite{Cui06,Roy11}. While the achieved shape accuracy and surface roughness is at a remarkable level \cite{Takahashi14,Ott16,Trichet15}, even nano-scale deviations of the surface from the ideal spherical shape affect the mode structure.
It has been shown that such imperfect mirror shape and finite mirror size can lead to loss, mode deformation, and shifted resonance frequencies at particular mirror separations \cite{Benedikter15}. The behavior is explained by resonant coupling between different transverse modes of the cavity and mode-dependent diffraction loss. The effects can be quantified by a model based on resonant state expansion \cite{Kleckner10} that takes the measured mirror profile into account.

Here, we show that mode coupling is also the origin of artefacts in SCM images due to imperfections of superpolished planar mirrors. In the first class of artefacts, the localized structures correspond to a spatially varying resonance condition for transverse mode coupling. It can be traced back to the surface topography of the planar mirror and its local curvature and gradient, which both lead to a spatial variation of the effective radius of curvature the mode samples, either on the nominally planar or on the concave micro-mirror. For the second class of artefacts, analysis of the background patterns' Fourier components reveals non-resonant admixture of specific higher-order modes to the fundamental cavity mode. Nano-scale roughness of the planar mirror leads to a spatial variation of the admixture which results in the observed high spatial frequency patterns.

These artefacts can have significant impact on SCM measurements, and we discuss ways to avoid the effects and also point towards their possibly beneficial use.

\begin{figure*}
	\centering
	\includegraphics[width=\textwidth]{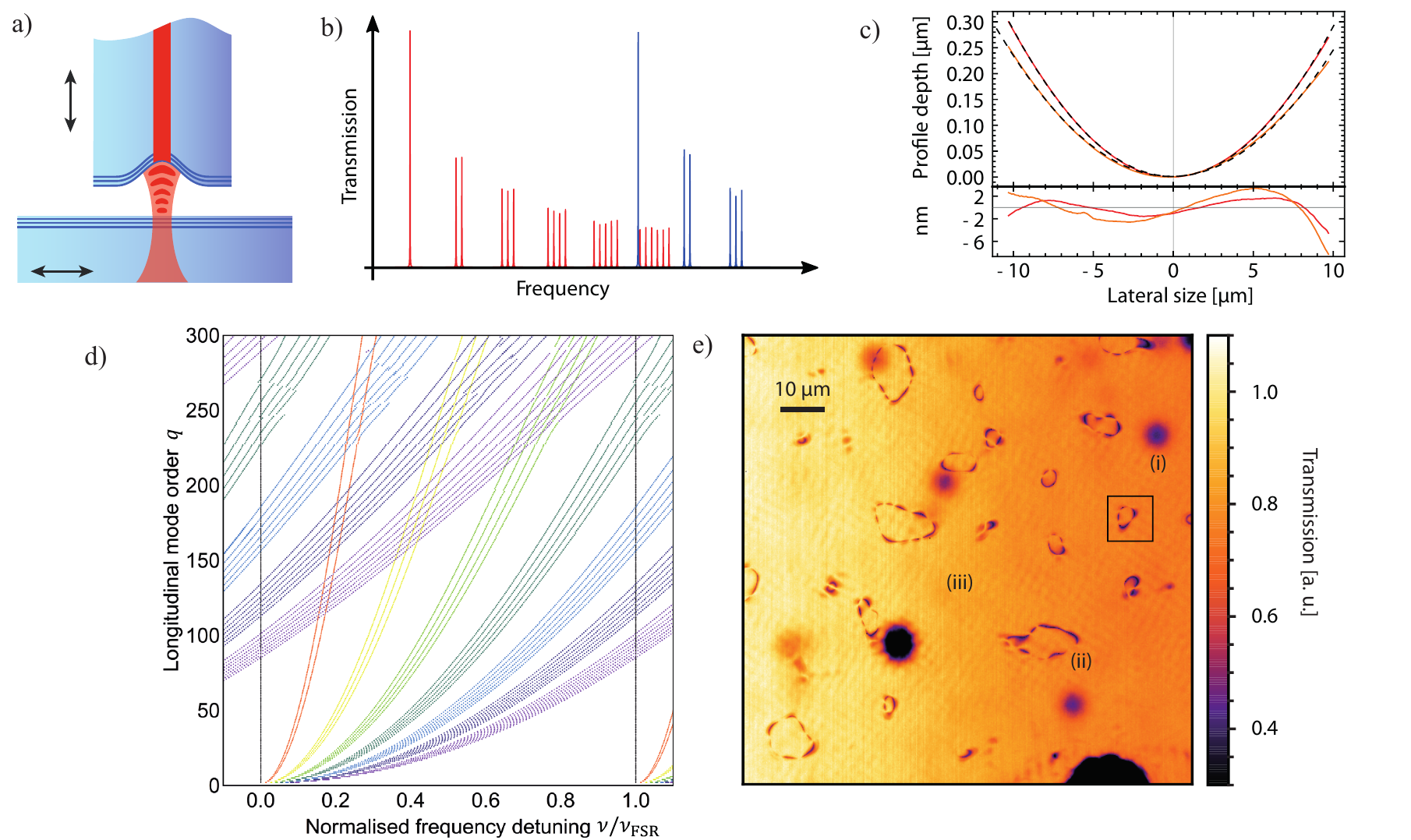}
	\caption{\label{fig1} 
		(a) Schematic scanning cavity setup showing the machined fiber, the plane mirror, and the optical mode.
		(b) Schematic cavity transmission spectrum showing how the fundamental mode (blue) can become resonant with a higher order transverse mode (red) of the neighboring longitudinal mode order.
		(c) Upper panel: Fiber profile section in $x$ and $y$ direction. Dashed lines: Parabolic fit. Lower panel: Residuals of the fits.
		(d) Simulated frequency detuning with respect to the fundamental mode (black horizontal lines) normalized to one free spectral range as a function of longitudinal mode order $ q $.Modes are shown for transversal mode orders $ \tilde{m}=1 $ (red), $ \tilde{m}=2 $ (yellow), ... , $ \tilde{m}=7 $ (purple).
		(e) Resonant cavity transmission scan of a $ 100 \times 100~\mathnormal{\mu m} $ area at $ q= 15$. (i) Loss introduced by a nano-scale particle yielding the Gaussian point spread function of the fundamental cavity mode. (ii) Contour line of resonant mode coupling. (iii) Periodic background pattern. Black square: scan area of Fig. \ref{fig3} (d)-(f).
	}
\end{figure*}

\section{Experimental setup and observations}
The cavity design studied here is depicted schematically in Fig. \ref{fig1} (a): the resonator consists of a macroscopic plane mirror and a concave micromirror on the end-facet of a single mode optical fiber machined with a CO$_2$ laser. The super-polished plane mirror substrate (fused silica, rms roughness $< 0.2$nm) is coated with a Bragg mirror for a center wavelength at 780~nm terminated with a $\lambda/4$ layer of SiO$_2$, yielding a transmission of $T_1=60$~ppm and a combined absorption and scattering loss  $L_1=20$~ppm. The fiber tip is coated for the same center wavelength with $T_2=12$~ppm and $L_2=12$~ppm. At short mirror separation, a cavity built from the two mirrors achieves a measured finesse of $\F= 60 000$, consistent with the above numbers. Both mirrors are characterized with white light interferometry to obtain the surface topography with a vertical resolution of $< 0.2~$nm and a diffraction limited lateral resolution of $~450~$nm. Figure \ref{fig1} (c) shows cuts through the measured fiber mirror profile along two orthogonal axes together with parabolic fits, yielding radii of curvature of $r_c^x=161~\mu$m and $r_c^y=201~\mu$m. The residual shows deviations on a few-nm scale, which are the origin of transverse-mode coupling effects.

The procedure described in \cite{Benedikter15} based on resonant state expansion \cite{Kleckner10} is used to calculate the mode frequencies $\nu_{qmn}$ of the studied cavity based on the measured surface profile, where $q$ is the longitudinal, and $m,n$ the transverse mode number of Hermite-Gaussian modes. Figure \ref{fig1}b) shows schematically the transverse mode spectrum for a given mirror separation, indicating the situation where a higher-order mode becomes (near-) degenerate with a fundamental mode of a neighboring longitudinal mode order. Figure \ref{fig1} (d) shows a calculation of the mode resonance spectra as a function of the mirror separation, where the free spectral range is normalized such that the fundamental modes appear at fixed frequencies. The different transverse-mode families $\tilde{m}=m+n$ are color coded, and their sub-modes are non-degenerate due to the presence of mirror ellipticity \cite{Uphoff15}. For a significant fraction of longitudinal modes, the fundamental mode becomes near-resonant with higher order modes.

In the experiment, the light of a grating-stabilized diode laser at 780~nm is coupled into the cavity through the fiber, and light transmitted through the plane mirror is detected with an avalanche photo diode. The laser frequency is fixed and we modulate the cavity length to record the resonance of the fundamental mode. We laterally raster-scan the planar mirror with a step size of 200~nm to record SCM images of the resonant cavity transmission for the fundamental mode at different longitudinal mode orders $q$. Figure \ref{fig1} (e) shows a measurement on a mostly clean mirror that shows three types of structures: (i) circularly symmetric Gaussian shaped features, (ii) often closed, sharp contour lines, and (iii) a periodic background pattern. The Gaussian shapes (i) originate from loss introduced by contaminants with a size much smaller than the cavity mode, such that the feature shape is given by the intensity distribution of the fundamental cavity mode as long as the introduced loss is small. This typically represents the signal of interest e.g. for cavity-enhanced extinction microscopy, and the mode size directly yields the spatial resolution of the measurement. Structures (ii) and (iii) are the focus of this study. It is apparent that they are much more localized than (i) and thus contain much higher spatial frequencies than the fundamental mode, such that they cannot originate from the fundamental mode alone.
We proceed by discussing the artefacts (ii) and (iii) in detail.

\begin{figure}
	\centering
	\includegraphics[width=\textwidth]{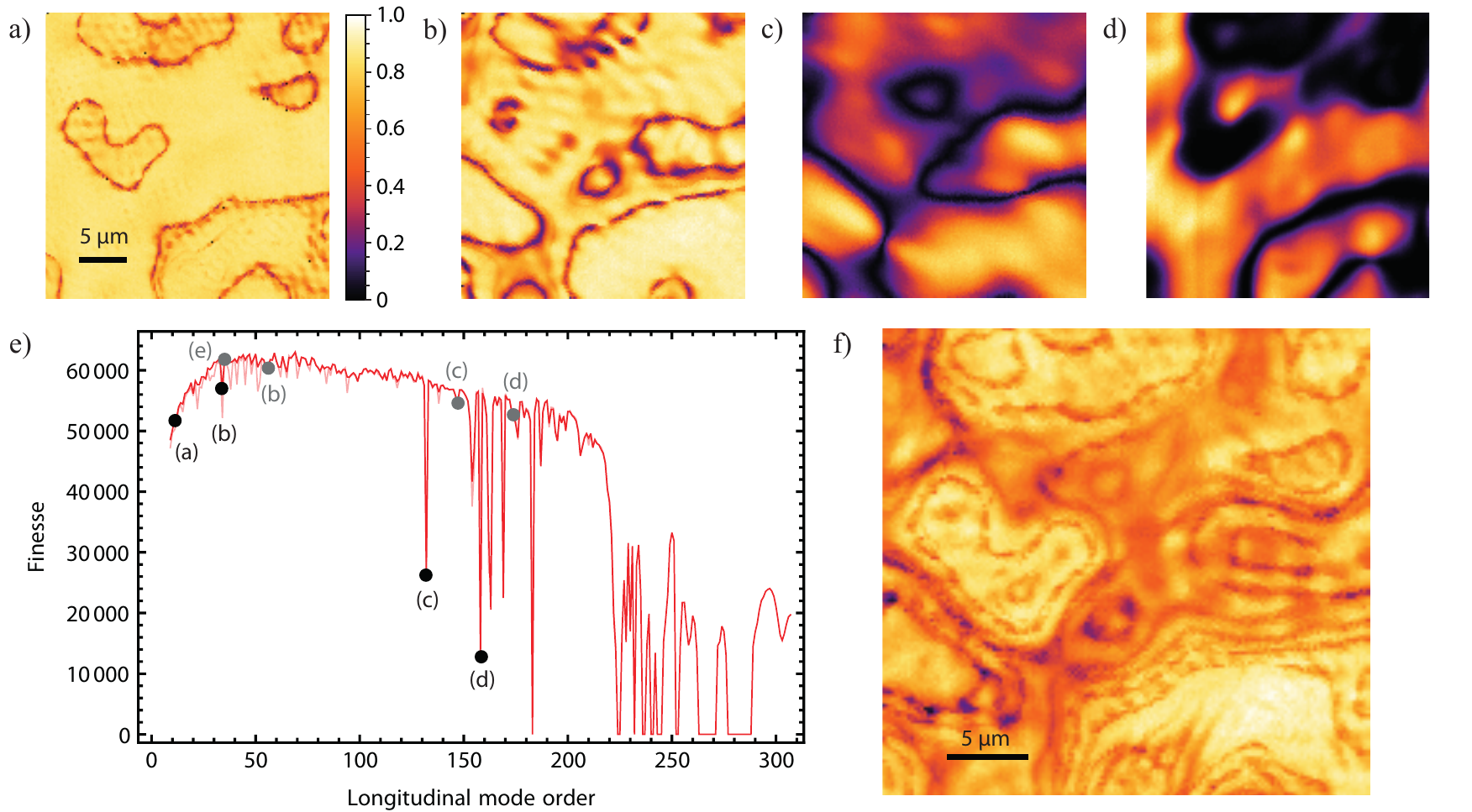}
	\caption{\label{fig2}
		(a) - (d) Resonant cavity transmission scans for mode orders $ q= 11,34,132,158$. The transmission is normalized to the peak value.
		(e) Cavity finesse as a function of longitudinal mode order, deduced from taking the predominant (solid red) and the average (light red) value of each transmission scan and converted to finesse knowing the mirror losses.  Black dots: mode orders of scans (a) to (d). Grey dots: mode orders of scans (b) to (e) in Fig. \ref{fig6}.
		(f) Overlay of transmission scans of longitudinal mode orders $ q= 11,12,14,16,22,24,28,38 $ showing isocontours of the penetration depth into the dielectric layer stack of the mirror. Arbitrary units. Color scale: see (a).
	}
\end{figure}

\section{Transverse-mode resonance isocontours}
To obtain a more detailed understanding of the properties of the contour lines (ii), we perform analogous measurements as shown in Figure~\ref{fig1} (e) on a defect-free, $ 30 \times 30 ~\mu$m area of a pristine planar mirror for all accessible longitudinal mode orders $q$ of the cavity. The complete data set is shown in the supplement. Figure~\ref{fig2} (e) shows the evaluated predominant (solid red) or the average (light red) finesse value  as a function of $q$ from this dataset. Specific mode orders show a reduced finesse, which in the latter case is related to the fraction of the area that is affected by the contour lines. Figures~\ref{fig2} (a-d) show example measurements at $q=(11,34,132,158)$, which are also indicated in Fig.~\ref{fig2} (e).

It can be observed that the shape of certain contours is very similar across different measurements, and that the width of the contour lines increases continuously with increasing $q$. Figure~\ref{fig2} (f) shows the superposition of measurements taken at mode orders $q=11,12,14,16,22,24,28,38$. Remarkably, the individual measurements combine to a consistent isocontour map.

\begin{figure}
	\centering
	\includegraphics[width=\textwidth]{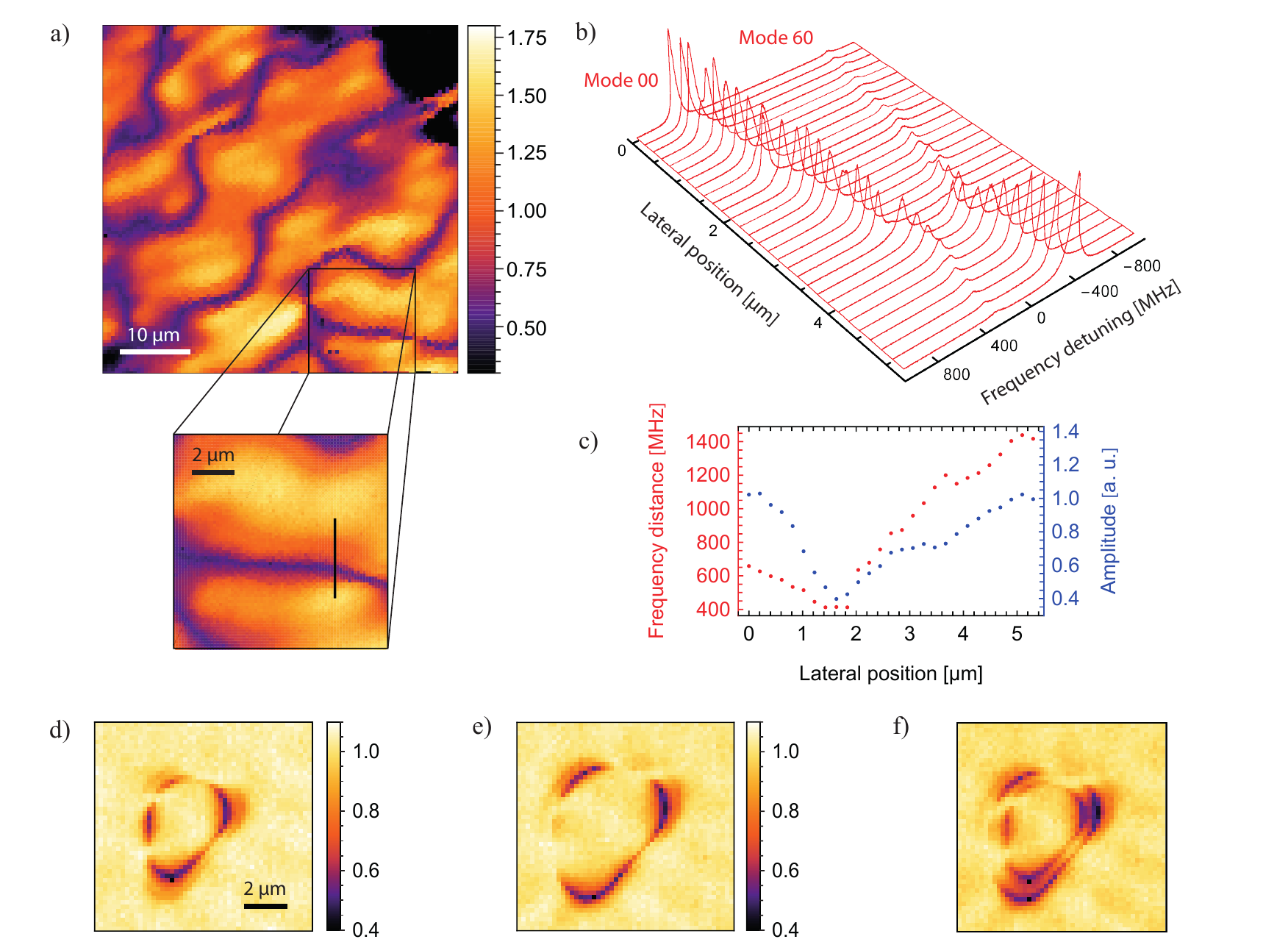}
	\caption{\label{fig3}
		(a) SCM transmission measurement at $ q=132 $  with a close up higher resolution scan. Pixel size: 0.5 and 0.2~\um. The black line indicates a $5.4~\mu$m long path along which spectra shown in (b) are taken.
		(b) Cavity transmission spectra around the fundamental mode for different lateral positions on the path in (a) showing an avoided crossing with mode $ (6,0) $. The minimal frequency separation is found at the darkest spot on the path in (a).
		(c) Maximal amplitude of the spectra in (b) (blue) and frequency separation $\Delta\nu_{00,60}$ of the two modes $ (0,0) $ and $ (6,0) $ along the path.
		(d),(e) Transmission scans at $ q=15 $. Changing the wavelength from $ \lambda=780.1400~$nm (d) to $ \lambda=780.2328~$nm (e) increases the size of the contour.
		(f) Scans (d) and (e) added up for comparison.}
\end{figure}

\subsection{Analysis}

A closer look at the cavity resonances in the surrounding of a contour line shows that mode degeneracy occurs at these spatial locations. In Figure~\ref{fig3} (a) we show an example transmission map taken at $q=132$. We follow the fundamental cavity resonance across a linear path perpendicular to a contour line, indicated in the inset. Figure~\ref{fig3} (b) shows cavity transmission spectra for each position on the mirror. We observe that in addition to the fundamental mode TEM$_{00}$, a higher order mode TEM$_{60}$ is visible, whose frequency detuning and strength changes with the lateral mirror position in the manner of an avoided level crossing. The frequency axis of the data is centered on the average frequency of the two modes to make this apparent. The higher order mode can be assigned unambiguously from the knowledge of $q$ and the mode frequency simulation shown in Figure~\ref{fig1}, as well as from counting the higher-order modes in the cavity transmission spectrum. At the location of the contour line, the mode separation is minimal, and the two modes have equal amplitude. Figure~\ref{fig3} (c) shows the quantitative evaluation of the data in (b), yielding a variation of the frequency separation $\Delta\nu_{00,60}\approx \pm 1~$GHz or equivalently $\pm20$ cavity linewidths. This evidences that the contour lines result from a spatially varying transverse-mode resonance, where the minimal mode separation on the contour line directly yields the inter-mode coupling strength.
This observation implies that the transverse modes experience different frequency shifts on different locations on the mirror.

The origin of the differential frequency shift can be constrained by considering the expression for the resonance frequency for a mode with longitudinal ($q$) and transverse $(m,n)$ order,
\begin{equation}\label{eq1}
\nu_{qmn}=\nu_\mathrm{FSR}\left(q+\frac{m+1/2}{\pi}\zeta^x+\frac{n+1/2}{\pi}\zeta^y\right),
\end{equation}
where $\nu_\mathrm{FSR}=c/(2d)$ is the free spectral range,
\begin{equation}
 \zeta^{x,y}=\arccos\sqrt{\left(1-\frac{d}{r_{c}^{x,y}}\right)\left(1-\frac{d}{r_{c,\mathrm{pl}}^{x,y}}\right)}
 \end{equation}
 the Gouy phase, $d=q\frac{\lambda}{2}+d_\mathrm{pen}$ is the effective cavity length, $r_c^{x,y}$ is the fiber mirror radius of curvature along the profile's major axes ($x,y$), $r_{c,\mathrm{pl}}^{x,y}$ the local radius of curvature of the nominally planar mirror, and $d_\mathrm{pen}=d_\mathrm{pen}^\mathrm{fiber}+d_\mathrm{pen}^\mathrm{plan}$ the optical field penetration depth into the dielectric mirror. For a quarter-wave Bragg mirror at the design wavelength as used for the fiber here, it is given by
\begin{equation}
d_{pen}^H=\frac{\lambda}{4\left(n_{H}-n_{L}\right)},
\end{equation}
while for the low-refractive index terminated planar mirror, the field penetration is larger according to \cite{Brovelli95}
\begin{equation}
d_{pen}^L=d_{pen}^H \bar{n}^2,
\end{equation}
with $\bar{n}=2(1/n_L+1/n_H)^{-1}$.
One can see that the mode frequency difference
\begin{equation}\label{eq4}
\Delta\nu=\nu_{qmn}-\nu_{(q+1)00} =\nu_\mathrm{FSR} \left[\frac{m}{\pi}\zeta^x+\frac{n}{\pi}\zeta^y -1\right]
\end{equation}
of a mode pair with a given set of $(q,m,n)$, can be varied by a change of $q$, the laser wavelength $\lambda$, the effective mirror radii of curvature $r_{c}^{x,y},r_{c,\mathrm{pl}}^{x,y}$ the modes experience \cite{Benedikter15}, and a change of the field penetration depth $d_{pen}$, which itself depends on the layer thickness and the refractive indices $n_H, n_L$.
%For the second equality in the above equation we have introduced the full width half maximum cavity linewidth $\kappa$ and simplified the expression by assuming a symmetric profile.
Note that the free spectral range $ \nu_{\mathrm{FSR}}=c/(2d) $ depends on the penetration depth, such that the given expression for $ \Delta\nu $ is an approximation in the case of varying $d_{pen}$.

\subsection{Spatial width of mode resonances}
The observed overall behavior of an increasing width of spatial mode resonances for increasing $q$ as seen in Fig.~\ref{fig2} (a-d) can be explained by the dispersion of the modes involved. From Equations~\ref{eq1}, \ref{eq4} and the condition of degeneracy, one can estimate the mode order leading to degeneracy at a given mirror separation by 
\begin{equation}\label{eq5}
\tilde{m}\approx\frac{\pi}{\mathrm{arccos}\sqrt{1-d/r_c}}.
\end{equation}
More accurately, the mode admixture to the fundamental mode can be calculated with the simulation of the full cavity mode structure \cite{Benedikter15}. Figure~\ref{fig6} (a) shows the contributions of transverse modes to the fundamental mode as a function of the longitudinal mode order. For short cavities, transverse modes with large $\tilde{m}$ become degenerate with the fundamental mode. These modes have a dispersion that differs largely from the fundamental mode and lead to a mode crossing under a large angle $\propto \tilde{m}$, see Fig.~\ref{fig1}(d). This means that small variations lead to a large frequency detuning between the modes, restricting resonant mixing to small spatial regions and thus the observed sharp features.
For long cavities, in contrast, the transverse-mode spacing is much larger, and modes with small $\tilde{m}$ become resonant with the fundamental mode. The difference between the modes' dispersion $\propto \tilde{m}$ is consequently smaller such that the resonant coupling
region is large, consistent with the much broader features observed for longer cavities.

\subsection{Wavelength dependence}
Next, we can directly evidence the dependence of the mode degeneracy location on $\lambda$ by taking SCM images at different laser wavelengths. Figure~\ref{fig3} (d,e) show two examples where we measure at $q=15$ and change the wavelength from $ \lambda=780.1400~$nm in (d) to $ \lambda=780.2328~$nm in (e), which changes the size of the mode degeneracy contour by about its spatial width. A wavelength change can thus globally change the observed contour pattern and could be used to map out contours across the entire mirror, but is not the origin of the structure. For all other measurements we have ensured a stable laser wavelength.

\subsection{Variation of local surface gradients}
\begin{figure}
	\centering
	\includegraphics[width=\textwidth]{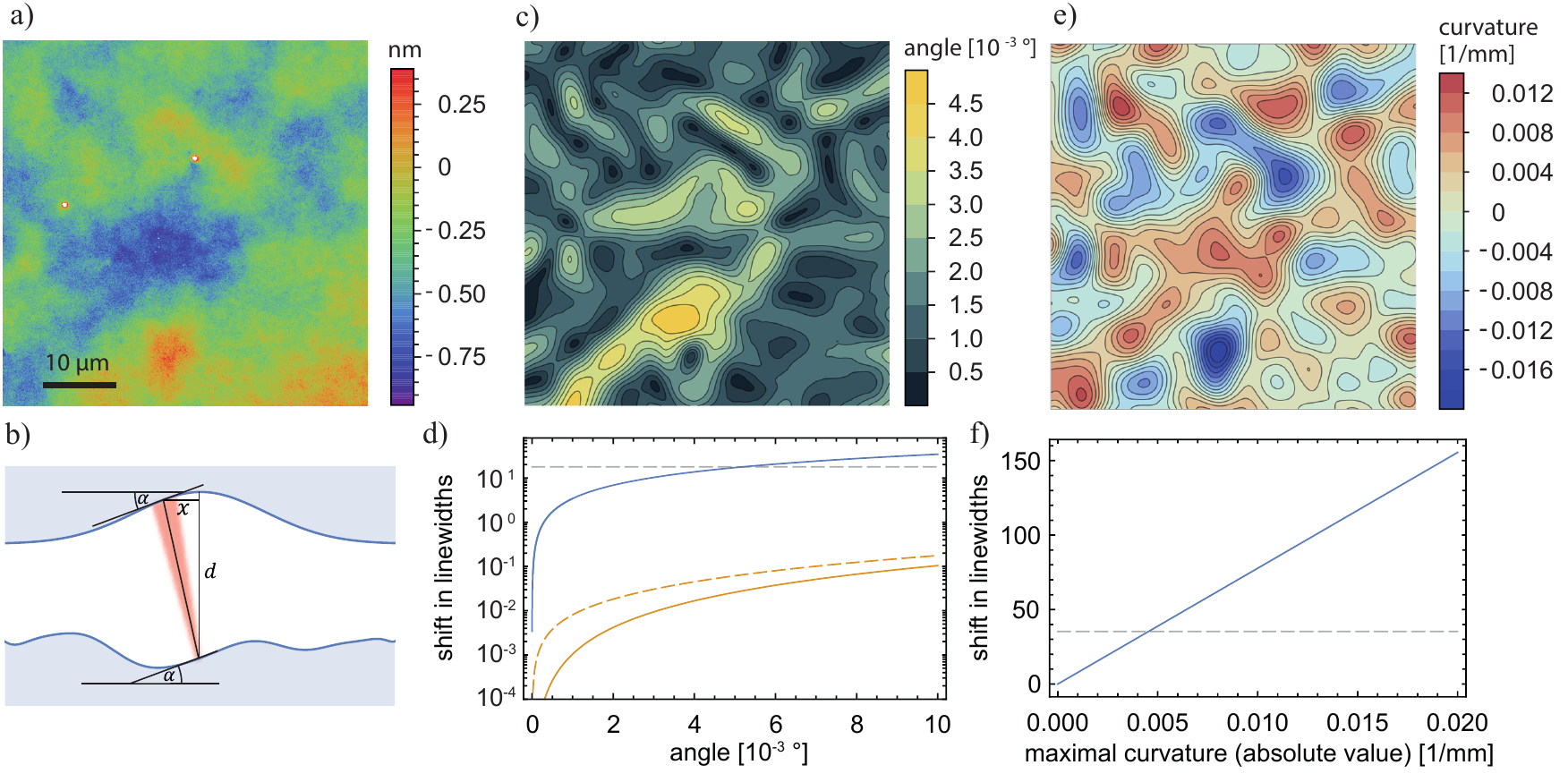}
	\caption{\label{fig4}
		(a) White light interferometric measurement of the mirror surface. (b) Schematic illustration of the effect of local gradients of the planar mirror. (c) Calculated modulus of the local surface angle and (e) average curvature of the area shown in (a).  (d) Change of $\Delta\nu_{00,60}$ as a function of surface angle at $q=132$ in units of $\kappa$. Yellow solid line: Frequency shift due to changing $r_c$ for an aligned cavity. Yellow dashed line: Shift due to changing $r_c$ and changing penetration depth. Blue line: Frequency shift due to $r_c$ for an initial misalignment of $\alpha_0=1.5^\circ$. Grey dashed line: Observed maximal shift. (f) Calculated change of transverse-mode frequency separation $\Delta\nu_{00,60}$ as a function of the maximal local curvature in units of $\kappa$ over the range of values observed in (a). Grey dashed line: Maximal shift observed in Fig.~\ref{fig3}.}
\end{figure}

We now consider a spatial change in the effective radius of curvature of the fiber mirror $r_c$ as a possible origin of the spatial variation of the mode degeneracy condition. Unevenness of the planar mirror leads to a spatially varying orientation of the cavity's optical axis, such that the cavity mode samples different areas on the micro mirror. Since the mirror shape is non-spherical, this translates into a change in $r_c$. Figure~\ref{fig4} (b) shows this situation schematically. To analyze the effect, we have taken interferometric images of the planar mirror topography on several locations to infer the typical variation of surface inclination. Figure~\ref{fig4}~(a) shows an example of a measured surface profile $h(x,y)$, and (c) shows the calculated modulus of the local surface angle $\alpha=\arctan(\sqrt{(\Delta h/\Delta x)^2+(\Delta h/\Delta y)^2})$ on the same area after smoothing the data with a Gaussian filter of a  $4.2~\mu$m $ 1/e^2 $ radius that matches the mode radius $w_0$ on the mirror. We observe typical height variations of up to $\Delta h\sim 1$~nm on length scales of $\Delta x\sim 10~\mu$m, yielding angles of up to $\alpha= 0.005^\circ$. Assuming that the cavity mode will form on the curved mirror where the two local surface angles match, we can calculate the change in the effective radius of curvature e.g. for a Gaussian profile $h(x)=-t\exp[-x^2/a^2]$ by solving for $x_\alpha$ for which $\theta=\arctan(\mathrm{d}h(x)/\mathrm{d}x)|_{x_\alpha}=\alpha$ and evaluating $r_c(x_\alpha)=[1+(\mathrm{d} h/\mathrm{d}x)^2]^{3/2}/(\mathrm{d}^2h/\mathrm{d}x^2)$. For a perfectly aligned cavity, this leads to relative changes in $r_c$ of $\Delta r_c/r_c\approx 10^{-6}$. Figure~\ref{fig4}~(d) shows a calculation of the resulting frequency shift $\Delta\nu_{00,60}$ as a function of $\alpha$ (yellow solid line), which is too small to explain the observed frequency shifts shown in Figure~\ref{fig3}~(c) (indicated by dashed gray line in Fig.~\ref{fig4}~(d)).
However, when the cavity is slightly misaligned, the change in $r_c$ with $x$ is much larger, and for an initial misalignment of $1.5^\circ$, the calculation shows that the observed range of frequency shifts is expected for the measured range of surface angles $\alpha$, see blue line in Fig.~\ref{fig4} (d). Misalignments of this order of magnitude are reasonable for the experiment reported here. This effect is thus a relevant contribution to the observations and can also be dominating.

\subsection{Variation of field penetration}
In principle, an additional contribution could arise from a spatial variation of the field penetration depth on the planar mirror. From the observed variation of $\Delta \nu$ we can quantify the necessary spatial variation of either the refractive index contrast or the layer thickness (or a combination of both). The observed  $\Delta \nu\approx \pm 1~$GHz corresponds to a change in the penetration depth of $28$~nm, which is about $3\%$ of the value for $d_{pen}^L$ of the planar mirror. This could be caused by a variation of the layer thickness by $3\%$, or alternatively by a change in the refractive index contrast by $7\times 10^{-3}$. Both values are at least an order of magnitude too large to be realistic.

We also consider a variation of field penetration at the fiber mirror due to local gradients at the planar mirror: Due to the partial directionality of the coating process \cite{Trichet15}, the coating layer thickness will depend on the local surface orientation, and the coating on the fiber will be thinner $\propto\cos(\theta+\theta_0)/\cos(\theta_0)$ at the inclined parts of the curved mirror. Here, $\theta_0$ is the deposition direction. For our fiber geometry, a maximal thickness variation up to $\pm3\%$ is expected for e.g.~a fully directional deposition under $\theta_0=30^\circ$. Figure~\ref{fig4}~(d) (yellow dashed line) shows that this effect is comparable in magnitude to the one due to the changing $r_c$ for a perfectly aligned cavity. However, the thickness variation is almost independent of $\alpha_0$ and thus does not contribute significantly for a tilted cavity.

\subsection{Variation of local curvature}
We now turn to the influence of the local curvature of the planar mirror due to surface imperfections. Figure~\ref{fig4} (e) shows the calculated average curvature on the area of Figure~\ref{fig4} (a). Therefore, we fit two-dimensional parabolas to each datapoint over an $8\mu$m area and take the average of the maximal and minimal curvature found along orthogonal eigenaxes. We find values that correspond to radii of curvature $r_{c,\mathrm{pl}}$ as small as 70~mm, which thus significantly affects the Gouy phase. The corresponding shift in mode frequency difference $\Delta\nu$ is shown in Figure~\ref{fig4} (f) as a function of maximal curvature. The values obtained from the surface curvature are comparable and even larger than the observed frequency variations. This suggests that this is a dominating effect in our experiments.

\begin{figure}
	\centering
	\includegraphics[width=\textwidth]{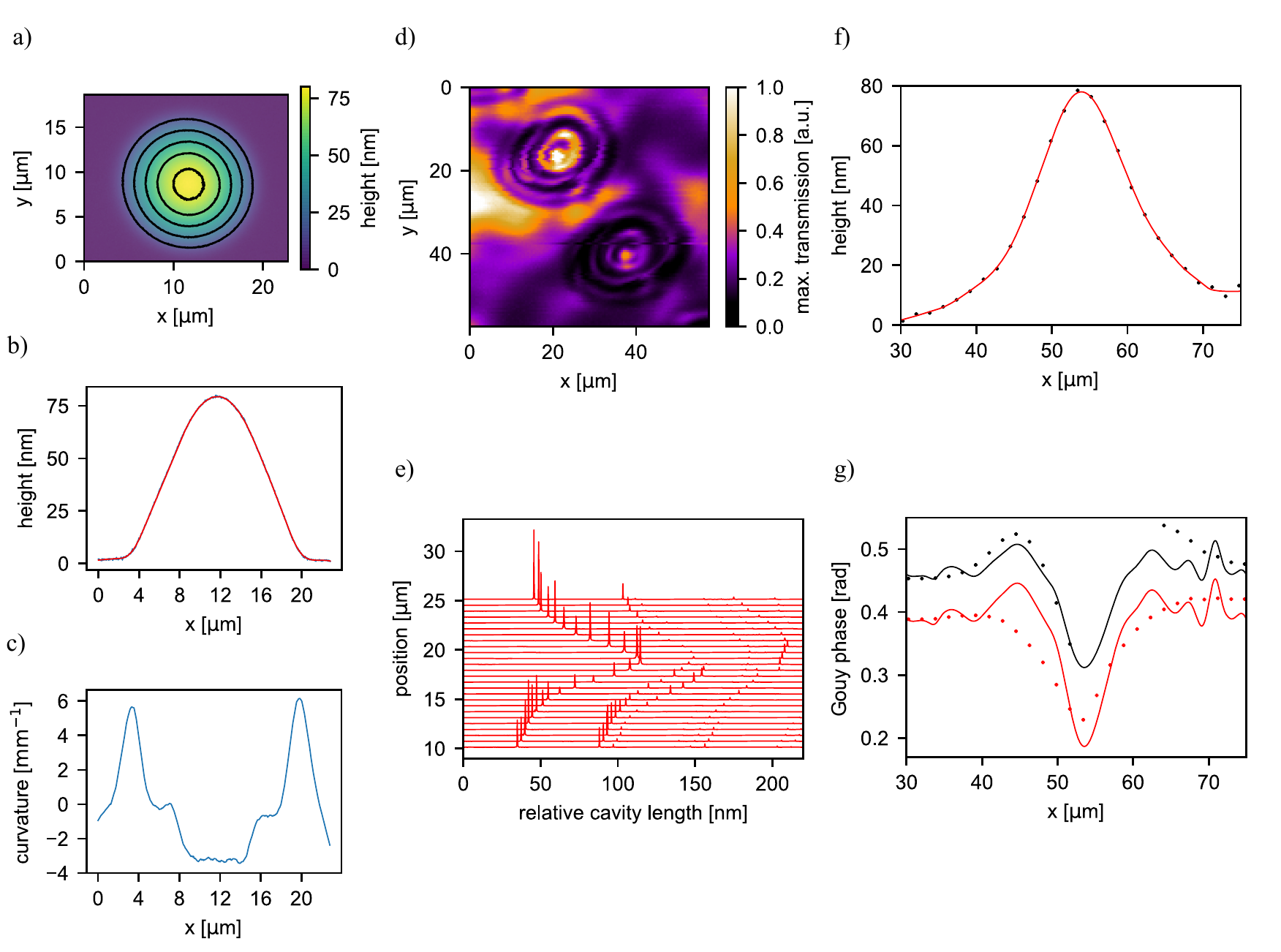}
	\caption{\label{fig5}
		(a) White light interferometric height measurement of a laser-induced buckle created on the planar mirror. (b) Cut through the data shown in (a) (black datapoints) together with a low-pass filtered copy. (c) Local curvature ($1/r_{c,\textrm{pl}}$) calculated from the low-pass filtered topography. (d) SCM measurement of an area with several buckles. (e) Cavity transmission as a function of the cavity length for different points across a buckle. (f) Height profile (black datapoints) evaluated from the position of the fundamental mode in (e). Red solid line: Interpolated and smoothed profile. (g) Measured Gouy phase difference to the fundamental mode for the TEM$_{01}$ (black datapoints) and TEM$_{01}$ mode (red datapoints). Solid lines: Calculated }
\end{figure}

To further support this picture, we have prepared a planar mirror with a microscopic surface profile that is much taller than the imperfections of the bare superpolished mirror, and that can be clearly identified with SCM measurements. Therefore we irradiate the mirror with a pulse of a focused CO$_2$ laser, leading to local delamination of the coating from the substrate and the formation of a dome. Figure~\ref{fig5} (a) shows a white light interferometric image of the structure together with a cut (b) and the calculated local curvature (c). The peak curvature is more than two orders of magnitude larger than on the pristine mirror, and its effect on the Gouy phase is thus unambiguously observable. We perform SCM measurements of such profiles, see Figure~\ref{fig5} (d), and observe circular isocontours of low cavity transmission as expected for mode mixing resonances. For these experiments, we use a different cavity operating at 860~nm with a fiber with $r_c^{x,y}=(280,295)~\mu$m. We use the measured relative cavity length at which the fundamental cavity mode appears to resolve the height profile of the dome. Figure~\ref{fig5} (e) shows transmission spectra taken along a linear path across one dome, and Figure~\ref{fig5} (f) shows the evaluated height. Compared to Figure~\ref{fig5} b) the profile is less sharp, consistent with the expected convolution of the cavity mode with the profile. In addition, we obtain the position of the modes TEM$_{01}$ and TEM$_{10}$ and calculate their repsective Gouy phase difference relative to the fundamental mode from the cavity length difference, see datapoints in Figure~\ref{fig5} (g). We observe an increase of $\zeta$ at the outer part of the profile and a reduction in the center, which fits well to the positive (concave) and negative (convex) curvature at the respective locations. In contrast, only a small influence of the surface gradient is observable, which would lead to an asymmetry of the Gouy phase profiles. We can also calculate the expected Gouy phase from the measured height profile and its curvature. The solid lines in Figure~\ref{fig5} (g) are the results of the calculation with $r_c^{x,y}$ taken as free fit parameters. We obtain $r_c^x=275~\mu$m and $r_c^y=380~\mu$m. While $r_c^x$ fits well to the measured value for this fiber, $r_c^y$ is larger, consistent with an angular misalignment of the fiber along this direction. This misalignment is also apparent from the different height of the TEM$_{01}$ and TEM$_{10}$ which originates from mode matching. The difference of the frequency variation of the two modes can be traced back to the orientation of the modes and the surface curvature. While the TEM$_{01}$ mode is oriented approximately along the scan direction and thus experiences both positive and negative curvature along its axis, the orthogonally oriented TEM$_{10}$ mode does to first order not experience a positive curvature and is only affected by the isotropic negative curvature at the dome's center.

Overall, the good matching of the measured and calculated Gouy phase across the dome confirms our interpretation, and we identify local curvature as the dominant effect also for pristine planar mirrors.

\section{Periodic background patterns}
The second type of artefacts observed in SCM measurements are periodic patterns in the resonant cavity transmission. In Fig.~\ref{fig1}(e), such a pattern (iii) is apparent over the entire measurement area. It contains higher spatial frequencies than contained by the fundamental mode, giving first evidence for the relevance of higher order transverse modes. Depending on $q$, the contrast of such patterns ranges between $<1\%$ and $30\%$, indicating that non-resonant coupling is present. This is further confirmed by the observed insensitivity of the pattern on wavelength changes. Even for a wavelength change of more than 30~nm, the pattern does not change significantly.

\begin{figure*}
	\centering
	\includegraphics[width=\textwidth]{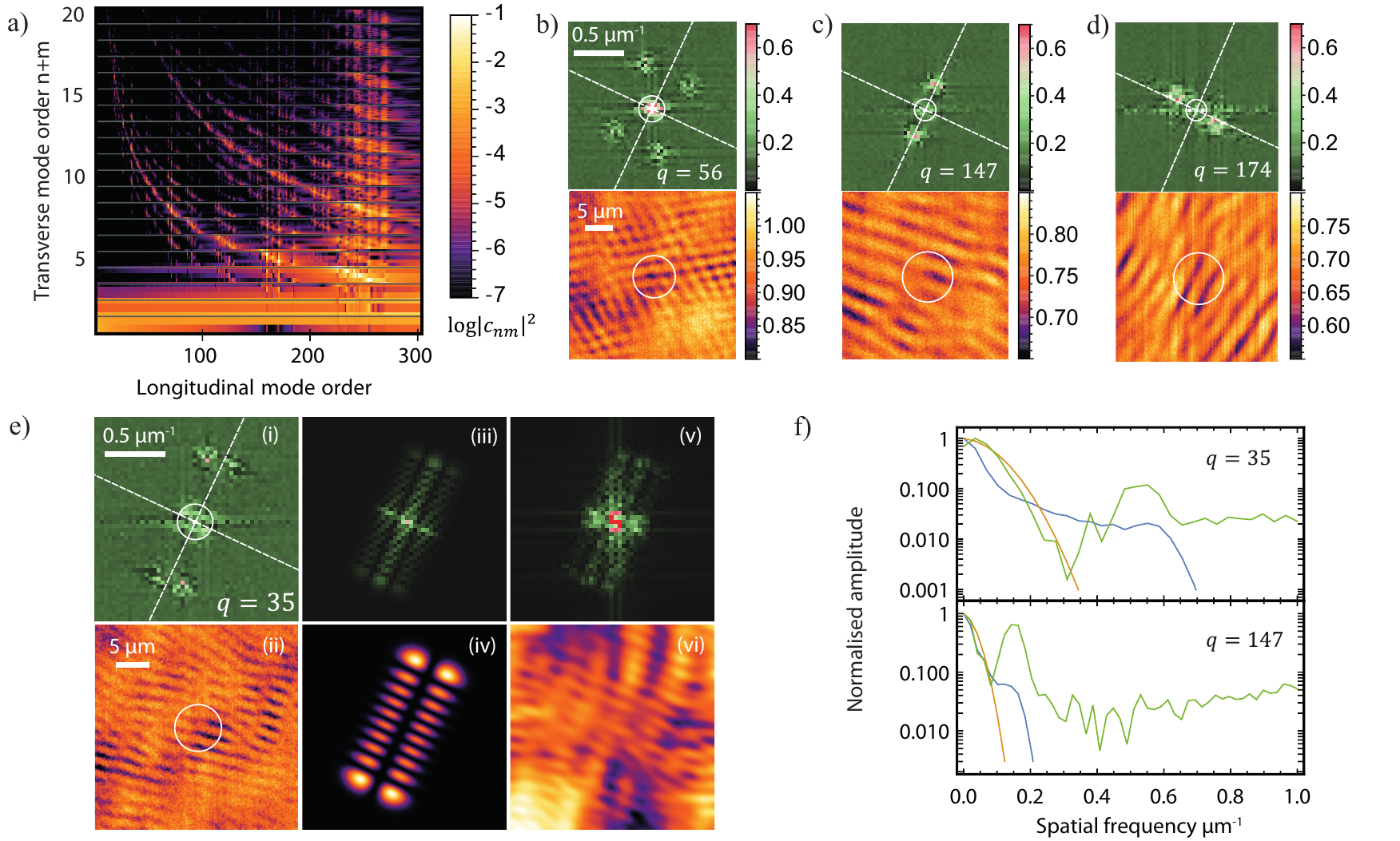}
	\caption{\label{fig6} 
		(a) Contributions $ c_{nm} $ of Hermite-Gaussian modes to the fundamental mode shown as $ \log \left|c_{nm}\right|^2 $ for different longitudinal mode orders. Each horizontal box separated by grey lines corresponds to $ m+n+1 $ modes of order $ m+n $ ranging from $ (0,m+n) $ to $ (m+n,0) $.
		(b) - (d) Lower panels: Transmission scans showing checker board and stripe background artefacts at different longitudinal mode orders. White circles indicate the $ 1/e^2 $-radii of the Gaussian modes at the respective cavity length. Upper panels: Spatial Fourier transforms. White circles: $ 1/e^2 $-radii of Fourier-transformed Gaussian modes. Dashed lines: Assumed principal axes of Hermite-Gaussian modes.
		(e) (i),(ii) Same as in (b) - (d) for $ q=35 $. (iv) H.-G. mode TEM$_{9,1} $ with its Fourier transform (iii). (vi) Convolution of normally distributed random numbers with mode TEM$_{9,1} $ and Fourier transform of the image (v).
		(f) Radial integration of Fourier transformed scans in (e) and (c) (green) together with the Gaussian mode (yellow) and a sum of modes of the order, which is contributing most to the ground mode: $ \sum_{i=0}^{\tilde{m}}(i,m+n-i) $ (blue), where $ \tilde{m}=10 $ for $ q=35 $ and $ \tilde{m}= 5$ for $ q=147 $.}
\end{figure*}

We investigate the background patterns for different $q$ and analyze their spatial frequency spectrum. Figures~\ref{fig6} (b)-(e) show SCM transmission measurements and corresponding two-dimensional (2D) spatial Fourier spectra. White circles indicate the spatial and spectral extent of the fundamental mode. The patterns show an orientation along the principal axes of the Hermite-Gaussian modes of the cavity, which are defined by the profile ellipticity of the fiber mirror. The Fourier spectra show localized maxima at high frequency, where the $q$ dependence of the observed frequency can be explained by the mode degeneracy condition approximated by eq.~\ref{eq5} and the respective spatial frequency spectrum of the dominant TEM$_{m,n}$ mode. A Hermite-Gaussian mode contains spatial frequencies up to $f_m=f_0\sqrt{m}$, where $f_0$ is the $1/e^2$ frequency of the fundamental mode. This results from the increase of the number of transverse field nodes $\propto m$ and the increase in mode size $\propto\sqrt{m+1}$.

Figure \ref{fig6} (f) shows a radial integration of the 2D Fourier spectrum and compares it with the spectrum of the Gaussian fundamental mode and the spectrum of the mode family that dominantly contributes to mode mixing as identified from the simulation shown in Fig.~\ref{fig6} (a). While the observed low frequency part matches well to the expected Gaussian mode spectrum for the particular longitudinal mode order, additional high frequency components agree with the frequencies of dominantly admixing modes. In particular, the localized peaks observed in the 2D Fourier spectra occur at the peak frequency of the most strongly admixing higher order mode. Components at higher frequency do not show any structure and can be attributed to uncorrelated measurement noise. 
%Interestingly, at frequencies between $f^{m,n}$ and $f_0$, the spectrum does not decay continuously, but shows smaller amplitudes. 
The overall scaling of observed and expected frequencies with $q$ can be clearly seen from the comparison of spectra for $q=35$ and $q=147$ (top and bottom panel of Fig.~\ref{fig6} f).

It is thus clear that a spatially varying non-resonant admixture of a few dominant higher-order modes leads to the observed patterns. We propose that the variation of the mode admixture originates from surface micro-roughness on the planar mirror. To support this hypothesis, we compare a measured structure at $q=35$ and its Fourier transform with the pattern formed by convoluting Gaussian white noise with the expected dominant admixing mode TEM$_{9,1}$ as obtained from the simulation shown in Figure~\ref{fig6} (a). Figure~\ref{fig6} (e) shows the measured transmission map, the TEM$_{9,1}$ mode, and the convolution of the TEM$_{9,1}$ mode with noise in the lower row, and the corresponding 2D Fourier spectra in the upper row. A similar texture with comparable length scales can be identified. However, the simulated Fourier spectrum shows more frequency components at intermediate frequencies $f_m > f>f_0$. The difference could be a sign for destructive interference with additionally contributing modes, which are indeed expected to be relevant from the simulation of admixing modes.

\section{Conclusion}
In summary, we have analyzed artefacts that regularly appear in scanning cavity microscopy and found consistent explanations that can describe the observations. We have identified two types of effects that can be traced back to variations of the local gradient and curvature of the surface topography that leads to resonant mode coupling, and to microscopic surface roughness which modulates the weak, non-resonant admixture of higher order modes. Both effects arise from transverse-mode coupling, which is significant due to imperfect shapes of both mirrors. In particular, even for superpolished mirrors, surface imperfections lead to a variation of the effective local radius of curvature that causes significant relative shifts of mode frequencies.%, reaching values of more than $1~$GHz in our experiments. %The amount of coupling can be quantified by observing avoided level crossings, and a coupling strength of $200$~MHz was found for one example shown here.

For quantitative measurements of e.g.\ the extinction contrast, the artefacts limit the achievable sensitivity \cite{Mader15}. The observed effects are also relevant for cavity QED experiments, in particular with solid state emitters. E.g.\ for the case of color centers in diamond, where thin diamond membranes are integrated into open-access cavities \cite{Janitz15,Riedel17,Bogdanovic17,Haeussler19}, mode coupling due to imperfect membrane surfaces can additionally affect the mode structure and limit the operation conditions for emitter-cavity coupling experiments.

 In general, this coupling can be reduced by minimizing deviations of the planar and curved mirror's surface profile from the respective ideal shape. Using e.g.\ multi-shot laser machining \cite{Ott16} or precise Focused-Ion-Beam milling \cite{Trichet15} potentially with CO$_2$ post-treatment \cite{Walker18}, concave profiles can be optimized to minimize mode coupling. In addition, longitudinal mode orders with weak mode coupling can be selected to fully avoid effect (ii) and largely suppress effect (iii). Also, our calculations indicate that for a well aligned cavity, the effect of local surface gradients on the planar mirror can be negligible. Finally, by using differential techniques e.g.\ with different wavelengths or with broadband illumination of the cavity, mode mixing can be subtracted or averaged out very efficiently.

In contrast, the amplifying effect of resonant mode coupling could be harnessed to sensitively characterize super mirrors for precision applications. E.g.\ smallest surface distortions and scattering could be analyzed with high spatial resolution, enabling new insight for challenging applications such as laser gyroscopes, high-finesse reference cavities, or gravitational wave detectors. 

\section*{Acknowledgments}

This work was partially funded by the European Union H2020 research and innovation programme under grant agreement No. 712721 (NanOQTech), and the DFG Cluster of Excellence Nanosystems Initiative Munich. We thank Theodor W. H{\"a}nsch for helpful discussions.\\

\end{document}